\documentclass[preprint,12pt]{elsarticle}
\usepackage{amssymb}
\usepackage{amsmath}
\usepackage{longtable}
\usepackage{array}
\usepackage{subcaption}
\usepackage{caption}
\usepackage{graphicx}
\usepackage{supertabular}
\usepackage[T1]{fontenc}
\usepackage[utf8]{inputenc}
\usepackage{lineno}
\usepackage[pdfencoding=auto]{hyperref}

\newcolumntype{x}[1]{>{\centering\arraybackslash\hspace{0pt}}p{#1}}
\usepackage[group-separator={,}]{siunitx}
\graphicspath{ {./Figures/} }
\journal{Astroparticle Physics}

\begin{document}
\begin{frontmatter}
\title
{Measurement of the Muon Flux at the Sanford Underground Research Facility with the LUX-ZEPLIN Dark Matter Detector}

\newcommand{\dagg}{\dagger}
\cortext[cor1]{Corresponding authors:}
\author{The LZ Collaboration}
\author[1,2]{D.S.~Akerib}
\author[3]{A.K.~Al Musalhi}
\author[3]{F.~Alder}
\author[4]{B.J.~Almquist}
\author[5]{C.S.~Amarasinghe}
\author[1,2]{A.~Ames}
\author[1,2]{T.J.~Anderson}
\author[6]{N.~Angelides}
\author[7,8]{H.M.~Ara\'{u}jo}
\author[9]{J.E.~Armstrong}
\author[1,2]{M.~Arthurs}
\author[10]{A.~Baker}
\author[8]{S.~Balashov}
\author[4]{J.~Bang}
\author[5]{J.W.~Bargemann}
\author[6]{E.E.~Barillier}
\author[14]{K.~Beattie}
\author[9]{A.~Bhatti}
\author[1,2]{T.P.~Biesiadzinski}
\author[6]{H.J.~Birch\corref{cor1}}
\ead{harveyjohnbirch@gmail.com}
\author[17]{E.~Bishop}
\author[15]{G.M.~Blockinger}
\author[8]{C.A.J.~Brew}
\author[11]{P.~Br\'{a}s}
\author[12]{S.~Burdin}
\author[13]{M.C.~Carmona-Benitez}
\author[12]{M.~Carter}
\author[16]{A.~Chawla}
\author[14]{H.~Chen}
\author[13]{Y.T.~Chin}
\author[20]{N.I.~Chott}
\author[18]{S.~Contreras}
\author[19]{M.V.~Converse}
\author[1,2]{R.~Coronel}
\author[3]{A.~Cottle}
\author[21]{G.~Cox}
\author[21]{D.~Curran}
\author[22,23]{C.E.~Dahl}
\author[3]{I.~Darlington}
\author[3]{S.~Dave}
\author[3]{A.~David}
\author[21]{J.~Delgaudio}
\author[24]{S.~Dey}
\author[13]{L.~de~Viveiros}
\author[7]{L.~Di Felice}
\author[4]{C.~Ding}
\author[10]{J.E.Y.~Dobson}
\author[19]{E.~Druszkiewicz}
\author[4]{S.~Dubey}
\author[21]{C.L.~Dunbar}
\author[25]{S.R.~Eriksen}
\author[24]{N.M.~Fearon}
\author[24]{N.~Fieldhouse}
\author[14]{S.~Fiorucci}
\author[25]{H.~Flaecher}
\author[12]{E.D.~Fraser}
\author[26]{T.M.A.~Fruth}
\author[1,2]{P.W.~Gaemers}
\author[4]{R.J.~Gaitskell}
\author[21]{A.~Geffre}
\author[13,20]{J.~Genovesi}
\author[3]{C.~Ghag}
\author[10]{J.~Ghamsari}
\author[15]{A.~Ghosh}
\author[14,27]{R.~Gibbons}
\author[28]{S.~Gokhale}
\author[3]{J.~Green}
\author[8]{M.G.D.van~der~Grinten}
\author[20]{J.J.~Haiston}
\author[9]{C.R.~Hall}
\author[12]{T.~Hall}
\author[6]{R.N~Hampp}
\author[29]{S.J.~Haselschwardt}
\author[6]{M.A.~Hernandez}
\author[30]{S.A.~Hertel}
\author[31]{G.J.~Homenides}
\author[21]{M.~Horn}
\author[18]{D.Q.~Huang}
\author[24,32]{D.~Hunt}
\author[3]{R.S.~James\fnref{fnRJ}}
\fntext[fnRJ]{Now at The University of Melbourne, School of Physics, Melbourne, VIC 3010, Australia.}
\author[7]{E.~Jacquet}
\author[11]{K.~Jenkins}
\author[16]{A.C.~Kaboth}
\author[18]{A.C.~Kamaha}
\author[15]{M.K.~Kannichankandy}
\author[19]{D.~Khaitan}
\author[8]{A.~Khazov}
\author[5]{J.~Kim}
\author[34]{Y.D.~Kim}
\author[14]{D.~Kodroff}
% ...

\address[1]{SLAC National Accelerator Laboratory, Menlo Park, CA 94025-7015, USA}
\address[2]{Kavli Institute for Particle Astrophysics and Cosmology, Stanford University, Stanford, CA 94305-4085 USA}
\address[3]{University College London (UCL), Department of Physics and Astronomy, London WC1E 6BT, UK}
\address[4]{Brown University, Department of Physics, Providence, RI 02912-9037, USA}
\address[5]{University of California, Santa Barbara, Department of Physics, Santa Barbara, CA 93106-9530, USA}
\address[6]{University of Zurich, Department of Physics, 8057 Zurich, Switzerland}
\address[7]{Imperial College London, Physics Department, Blackett Laboratory, London SW7 2AZ, UK}
\address[8]{STFC Rutherford Appleton Laboratory (RAL), Didcot, OX11 0QX, UK}
\address[9]{University of Maryland, Department of Physics, College Park, MD 20742-4111, USA}
\address[10]{King's College London, King’s College London, Department of Physics, London WC2R 2LS, UK}
\address[11]{Laborat\'orio de Instrumenta\c c\~ao e F\'isica Experimental de Part\'iculas (LIP), University of Coimbra, P-3004 516 Coimbra, Portugal}
\address[12]{University of Liverpool, Department of Physics, Liverpool L69 7ZE, UK}
\address[13]{Pennsylvania State University, Department of Physics, University Park, PA 16802-6300, USA}
\address[14]{Lawrence Berkeley National Laboratory (LBNL), Berkeley, CA 94720-8099, USA}
\address[15]{University at Albany (SUNY), Department of Physics, Albany, NY 12222-0100, USA}
\address[16]{Royal Holloway, University of London, Department of Physics, Egham, TW20 0EX, UK}
\address[17]{University of Edinburgh, SUPA, School of Physics and Astronomy, Edinburgh EH9 3FD, UK}
\address[18]{University of California, Los Angeles, Department of Physics \& Astronomy, Los Angeles, CA 90095-1547}
\address[19]{University of Rochester, Department of Physics and Astronomy, Rochester, NY 14627-0171, USA}
\address[20]{South Dakota School of Mines and Technology, Rapid City, SD 57701-3901, USA}
\address[21]{South Dakota Science and Technology Authority (SDSTA), Sanford Underground Research Facility, Lead, SD 57754-1700, USA}
\address[22]{Northwestern University, Department of Physics \& Astronomy, Evanston, IL 60208-3112, USA}
\address[23]{Fermi National Accelerator Laboratory (FNAL), Batavia, IL 60510-5011, USA}
\address[24]{University of Oxford, Department of Physics, Oxford OX1 3RH, UK}
\address[25]{University of Bristol, H.H. Wills Physics Laboratory, Bristol, BS8 1TL, UK}
\address[26]{The University of Sydney, School of Physics, Physics Road, Camperdown, Sydney, NSW 2006, Australia}
\address[27]{University of California, Berkeley, Department of Physics, Berkeley, CA 94720-7300, USA}
\address[28]{Brookhaven National Laboratory (BNL), Upton, NY 11973-5000, USA}
\address[29]{University of Michigan, Randall Laboratory of Physics, Ann Arbor, MI 48109-1040, USA}
\address[30]{University of Massachusetts, Department of Physics, Amherst, MA 01003-9337, USA}
\address[31]{University of Alabama, Department of Physics \& Astronomy, Tuscaloosa, AL 34587-0324, USA}
\address[32]{University of Texas at Austin, Department of Physics, Austin, TX 78712-1192, USA}
% ...
\address[34]{IBS Center for Underground Physics (CUP), Yuseong-gu, Daejeon, Korea}
% ...

\begin{abstract}
High-energy cosmic-ray muons reaching deep underground laboratories can cause events in detectors that mimic signals expected from dark matter particles, neutrinos, or rare decays. Knowledge of the muon flux and energy spectrum is important for evaluating the background rate caused by muons and their secondaries. In this paper, we report the measurement of the cosmic-ray muon flux in the Davis Campus of the Sanford Underground Research Facility with the LUX-ZEPLIN detector. Using 366.4~days of exposure, the muon rate through the detector was measured as $10.94\pm0.17_\textrm{stat.}~\textrm{day}^{-1}$ with energy thresholds of 20~MeV in the inner xenon detector and 8 MeV in the outer liquid scintillator detector. This rate corresponds to a muon flux of $(5.09\pm0.08_\textrm{stat.}\pm0.10_\textrm{sys.})\times10^{-9}~\textrm{cm}^{-2}\textrm{s}^{-1}$ in the Davis Cavern.
\end{abstract}

\begin{keyword}
Dark Matter Experiment,
Cosmic-Ray Muons, 
Underground Detectors
\end{keyword}

\end{frontmatter}

\section{Introduction}
The LUX-ZEPLIN (LZ) experiment is searching for Weakly Interacting Massive Particle (WIMP) dark matter in a wide range of WIMP masses and other rare phenomena~\cite{LZfirstResults,WS2024,LZ:2025igz,LZ:2025zpw,LZ:2024psa}. The detector houses a dual-phase xenon time projection chamber (TPC) with arrays of photomultiplier tubes (PMTs) at the top and bottom. The TPC is surrounded by a liquid xenon (LXe) Skin and an Outer Detector (OD) filled with a gadolinium-loaded liquid scintillator, both serving as anti-coincidence systems (veto detectors) to reject background events. The LZ experiment is operating in the Davis Campus of the Sanford Underground Research Facility (SURF) at the 4850~ft level. A complete description of the detector design is given in Ref.~\cite{LZ}.

Muon-related backgrounds pose a threat to high-sensitivity rare-event search experiments through various processes such as nuclear recoils from neutron elastic scattering (WIMP searches); neutron inelastic scattering and capture (neutrino detection, neutrinoless double beta decay, nucleon decays and other experiments).

This background is relatively small in existing experiments, but can potentially limit the sensitivity of larger detectors discussed or planned for construction at SURF, such as DUNE~\cite{DUNE:2020lwj}. Hence, muon measurements have a wider impact on science at SURF.

Previously, the muon intensity in the vertical direction (within $18^{\circ}$ from vertical) at SURF, in the same cavern as LZ (vertical depth of about 1480 m), was measured with the veto system of the Davis solar neutrino experiment to be $(5.38\pm0.07)\times10^{-9}~\textrm{cm}^{-2}\textrm{s}^{-1}\textrm{sr}^{-1}$\cite{cherry}.

We have included single and multiple muons in the value above by increasing the vertical intensity of single muons reported in Ref.~\cite{cherry} by the measured fraction of multiple muons from the same experiment (with a multiple-to-single ratio of 9.7\%).
The total muon flux has recently been measured by the veto system of the M\footnotesize{AJORANA} \normalsize{D}\footnotesize{EMONSTRATOR}\normalsize, located at the same level at SURF in the nearby cavern, as $(5.31\pm0.17)\times10^{-9}~\textrm{cm}^{-2}\textrm{s}^{-1}$~\cite{majorana}. 

The two fluxes above cannot be directly compared since the first one is reported for the vertical direction, while the second one is effectively integrated over all angles. 
Using a muon angular distribution predicted by MUSIC and MUSUN simulations (described in Section~\ref{sec:muon_sims}) and extrapolating the results from Ref.~\cite{cherry} to the total flux, the measurements from Ref.~\cite{cherry} are higher than those reported by M\footnotesize{AJORANA} \normalsize{D}\footnotesize{EMONSTRATOR }\normalsize by about 20\%. 
The difference may be caused, at least partly, by different locations of the two detectors (the distance is about 50~m between them), but a new measurement with the LZ in the Davis cavern should resolve the difference.

Cosmic-ray muon background in the LZ experiment was extensively studied based on a Monte Carlo simulation campaign, and the cosmogenic neutron background was evaluated to be negligible thanks to the high background rejection power, including veto detectors~\cite{LZ_SIMS}.

In this paper, we report the measurements and simulations of the muon event rate in LZ. The result helps tune the LZ muon model and provides a new measurement of the muon flux in the Davis cavern. Compared to previous measurements \cite{cherry,majorana}, we improve on statistical and systematic uncertainties. We have also evaluated with better precision the mean density of rock above the Davis campus, an important parameter for future muon simulations.

Section~\ref{sec:muon_sims} of the paper introduces the muon simulations. The selection of muon events is described in Section~\ref{sec:evt_selec}. The results of the muon rate measurement are presented in Section~\ref{sec:rate_meas}. The procedure of determining the muon flux from the measured rate is explained in Section~\ref{sec:eval_flux} together with the evaluation of the average rock density. Conclusions are given in Section~\ref{sec:conc}. 

\section{Muon Simulations}\label{sec:muon_sims}
The muon model presented below describes the muon energy spectrum and angular distribution. It is used together with the LZ measurement that constrains the total muon rate in the detector, to extract the muon flux at the Davis cavern.

The energy spectrum and angular distribution of muons in the Davis cavern were obtained using two Monte-Carlo codes: MUSIC (MUon SImulation Code) and MUSUN (MUon Simulations UNderground) \cite{music,musun}, adapted for the LZ experiment. MUSIC propagates muons with different energies from the Earth's surface through various thicknesses of rock, recording their energy spectra and angular distribution. 

The rock composition was derived from the measurements reported in Ref.~\cite{mei}, with additional input from Ref.~\cite{zhang}, giving average parameters of $\langle Z\rangle = 12.1$ and $\langle A\rangle = 24.2$. The average rock density was assumed to be $2.70~\textrm{g~cm}^{-3}$~\cite{zhang}. In the MUSIC simulations, the known surface profile was included~\cite{LZ_SIMS}, and the rock was assumed to have uniform composition and density above the detector.

Inside the LZ simulation framework BACCARAT \cite{LZ_SIMS} (based on the GEANT4 toolkit \cite{Geant4}), MUSUN reads MUSIC outputs to generate muons in the rock on the top and sides of a box that encompasses the Davis cavern at SURF. A number of physics lists are used by BACCARAT, the prominent modules used to model muon interactions with the detector are {\fontfamily{qcr}\selectfont EMLivermorePhysics} and {\fontfamily{qcr}\selectfont QGSP\_BIC\_HP\_Gd}. A full description of BACCARAT can be found in Ref.~\cite{LZ_SIMS}.

\begin{figure}[!ht]
\centering
    \begin{subfigure}{0.75\textwidth}
        \centering
        \centering\includegraphics[width=1\textwidth]{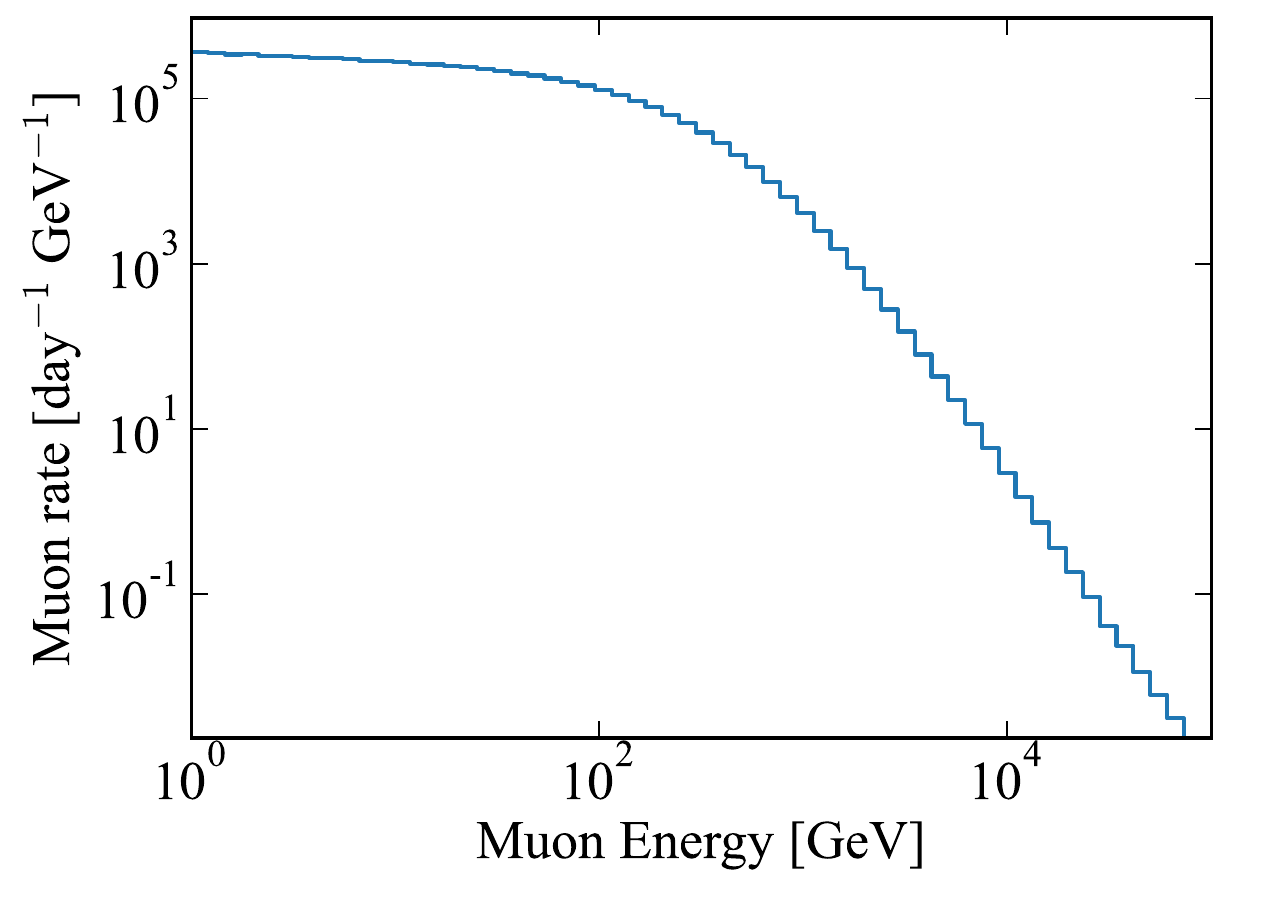}
        \caption{}
        \label{fig:Prim_E}
    \end{subfigure}
    \hfill
    \begin{subfigure}{0.75\textwidth}  
        \centering 
        \centering\includegraphics[width=1\textwidth]{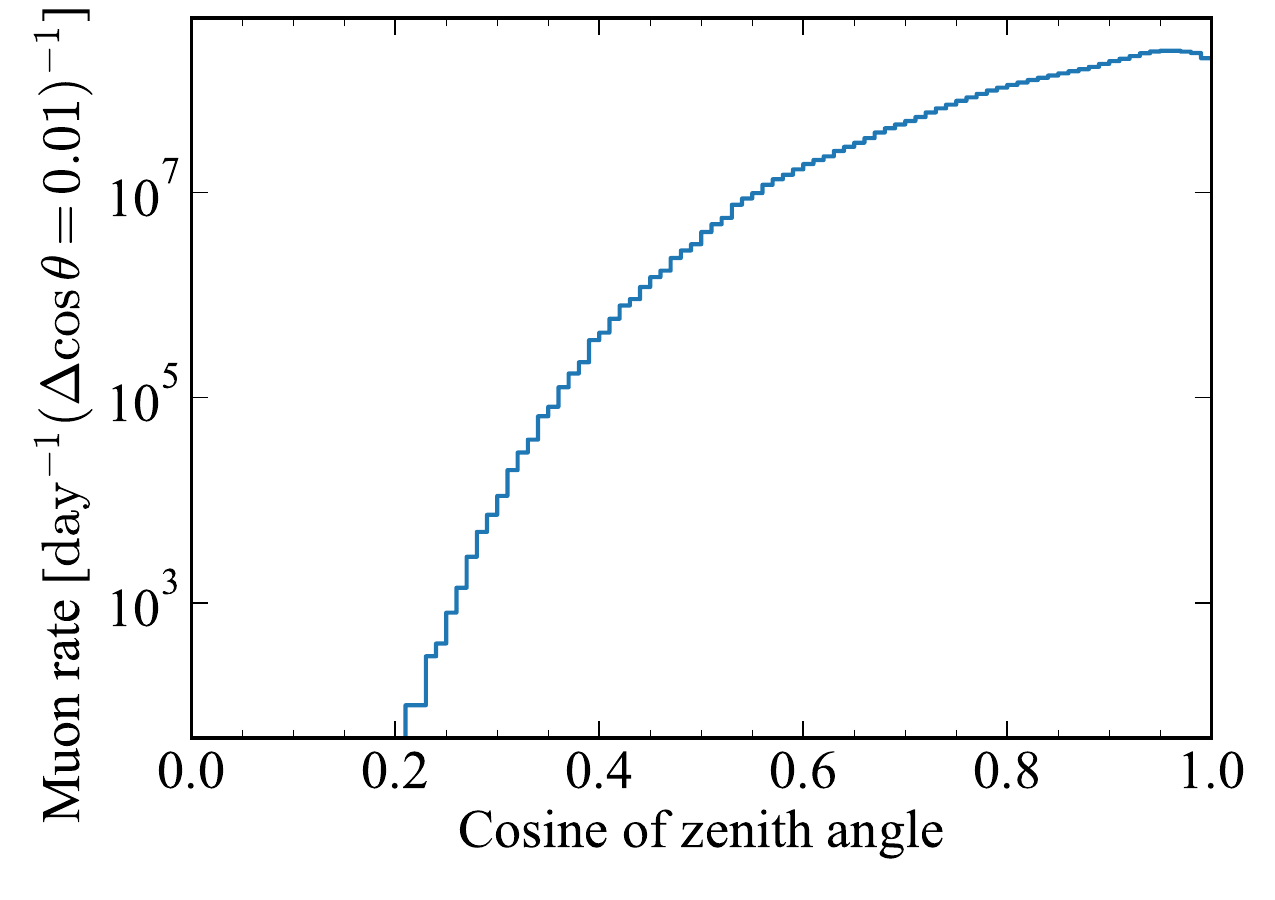}
        \caption{} 
        \label{fig:cosZ}
    \end{subfigure}
    \caption{Energy spectrum (a) and zenith angular distribution (b) of simulated muons at SURF's Davis cavern, generated with the MUSUN model at the surface of the box enclosing the cavern boundary (see text for details).}
    \label{fig:Prime_info}
\end{figure}

The ratio of $\mu{^+}/\mu^{-}$ is set to 1.38 based on measurements of high-energy muons \cite{Ashley:1975uj}. The box extends from 7~m above the cavern boundary to 3~m below. The vertical walls of the box are positioned 5~m away from the boundary. Moving the box surfaces into the rock, away from the cavern, ensures the development of muon-induced cascades in the rock before muons and secondary particles produced by muons in rock enter the cavern, which is important when studying muon-induced backgrounds. Cascades starting outside the box are not expected to reach the cavern. The energy spectrum and the zenith angular distribution of the muons generated by MUSUN on the surface of the box are shown in \autoref{fig:Prime_info}.

This muon model has been described in detail in Ref.~\cite{LZ_SIMS}, and has been used to calculate muon fluxes and estimate the expected muon-induced background in LZ. 
We propagated $4.82\times10^7$ muons through the box in the simulation and recorded muon energy depositions in the OD, Skin, and TPC.

This corresponds to 9147~days of exposure (0.0610 muons per second through the box), a factor of 25 more than for data collected during the WIMP searches conducted between December 23, 2021, to May 11, 2022 (WS2022~\cite{LZfirstResults}) and March 27, 2023, to April 1, 2024 (WS2024~\cite{WS2024}).

Our calculations within this model predict a vertical muon intensity of $5.18\times10^{-9}~\textrm{cm}^{-2}\textrm{s}^{-1}\textrm{sr}^{-1}$, within 3 standard deviations from the measured value of $(5.38\pm0.07_\textrm{stat.})\times10^{-9}~\textrm{cm}^{-2}\textrm{s}^{-1}\textrm{sr}^{-1}$ in the Davis experiment~\cite{cherry}.
The total muon flux through a spherical detector with unit cross-sectional area has been calculated as $6.16\times10^{-9}\textrm{cm}^{-2}\textrm{s}^{-1}$, which is 16\% higher than the recently measured value of $(5.31\pm0.17_\textrm{stat.})\times10^{-9}~\textrm{cm}^{-2}\textrm{s}^{-1}$ by the M\footnotesize{AJORANA} \normalsize{D}\footnotesize{EMONSTRATOR }\normalsize~\cite{majorana}.

The difference between the two measurements and our initial model indicates a possible uncertainty in the model of about 20\%. The measurement will help tune the LZ muon model, in which the assumed rock density is the main source of systematic uncertainty.

\section{Muon Event Selection}\label{sec:evt_selec}
\begin{figure}[!ht]
    \centering
    \includegraphics[width=\textwidth]{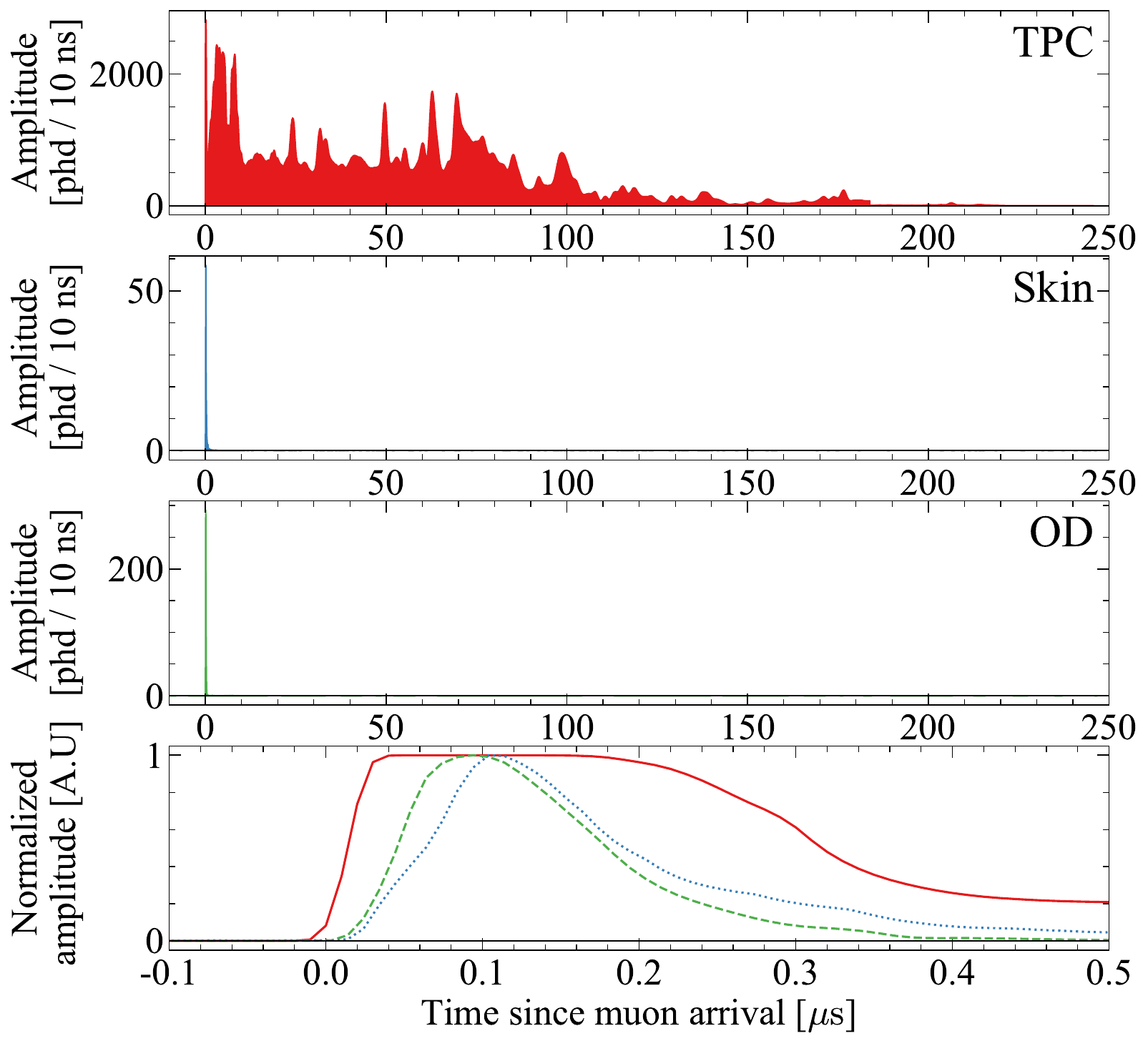}
    \caption{Waveform output of a typical muon event captured by the LZ data acquisition system. The amplitude is measured in single photons registered by PMTs per 10~ns sample. The individual waveform traces from the TPC, Skin, and OD are shown in the top three panels. The bottom panel shows a magnified view of the pulses in all detectors overlaid at the time the muon initially passes through the experiment to demonstrate the contribution of pulse timing to the muon event selection. The TPC pulse corresponds to the solid red line, the OD pulse corresponds to the dashed green line, and the Skin pulse corresponds to the dotted blue line.}
    \label{fig:eventViewer}
\end{figure}

Recorded waveforms of muon events have distinct features that have been used to select these events to evaluate the muon rate in LZ. Two datasets, WS2022 and WS2024, with operating conditions described in Refs.~\cite{LZfirstResults,WS2024} and detector live time of 366.4 days have been considered in the muon analysis.
Cosmic-ray muon events in the LZ experiment can be uniquely identified by their large energy deposits that occur nearly simultaneously across the three detector systems: OD, Skin, and TPC.
In the OD and Skin, a muon produces a singular pulse at time $t_{\textrm{OD}}$ and $t_{\textrm{Skin}}$, respectively. The muon signal in the TPC constitutes a series of consecutive pulses from the ionisation of atoms along the muon track. At every interaction point along the track, there will be a primary scintillation pulse (`S1') and an ionisation pulse (`S2'). The S2 pulse is caused by ionisation electrons drifting from the interaction site in the liquid volume to the gaseous phase at the top of the TPC, producing electroluminescence~\cite{LZ}. Given the very short time the muon takes to cross the TPC (a few nanoseconds) all of the scintillation signals form a single S1 pulse. 
The S2 signals merge to create a long, continuous pulse as the muon ionises atoms along its path, as illustrated in \autoref{fig:eventViewer}. If the muon enters the TPC from the top, the S2 pulses begin immediately after the S1 pulse.
When a muon passes through the TPC, the long duration of the S2 signals (widths of up to the maximum TPC drift time, compared to the few microseconds widths observed during regular operation) fill the memory buffers used to store the waveforms before the end of the event window. This results in the loss of signals from $\sim180~$\textmu s until the end of the event. An example of the ``event clipping'' caused by buffer filling can be seen in \autoref{fig:eventViewer}. The start of the first pulse in the TPC is used in the inter-detector timing selection as $t_{\textrm{TPC}}$. 

Events that exhibited the coincidence between the TPC, Skin and the OD were initially identified, and through the analysis of the inter-detector timing spectra, the three different criteria were chosen as follows:
 \begin{enumerate}
    \item $-200~\textrm{ns} < \Delta t_{\text{OD - Skin}} < 200~\textrm{ns}$,
    \item $-1200~\textrm{ns} < \Delta t_{\text{OD - TPC}} < 200~\textrm{ns}$,
    \item $-1200~\textrm{ns} < \Delta t_{\text{Skin - TPC}} < 200~\textrm{ns}$.
\end{enumerate}
Here $\Delta t_{X-Y}=t_X-t_Y$ denotes the difference in time between the pulses observed in detectors $X$ and $Y$ with $t_{X,Y}$ being the time at which the pulse reaches 5\% of its total pulse area.   
Criterion~1 is defined as the time difference between the largest pulses in the OD and the Skin. Criterion~2 is the time difference between the largest pulse in the OD and the start of the muon tail in the TPC. Criterion~3 is the time difference between the largest pulse in the Skin and the start of the muon tail in the TPC. Events are excluded from the muon event selection if they fail to satisfy all three criteria.

Background events, such as those induced by high-energy gamma-rays from radioactivity or caused by neutron capture on gadolinium in liquid scintillator, can trigger all three detectors. Therefore, an OD threshold of 2000~phd (photons detected) corresponding to about 8~MeV energy deposition in the OD~\cite{OD_calib} was used in the data analysis. 

No explicit energy threshold, other than the 0.1~phd requirement to trigger the data acquisition system \cite{LZ:2024bvw}, is applied to the signals in the xenon Skin due to a highly non-uniform response of the Skin to an energy deposition.

\section{Muon Rate Measurement}\label{sec:rate_meas}
The same muon event selection cuts have been applied to both LZ data and simulations so the measured and simulated muon event rates above a chosen TPC energy threshold can be compared, and the model normalisation factor can be computed. Determining the threshold for energy deposited in the TPC is a critical part of obtaining an accurate muon rate. A particular concern is the effect of muon-induced secondary particles, which may not be modelled to a high degree of precision. Low-energy deposits in the TPC may be caused by events where a muon passed only through the OD and its secondaries caused coincident signals in the Skin and TPC. 
The threshold for the energy deposits in the TPC should be optimised to mitigate this uncertainty, such that a robust scaling factor between the simulated event rate and the measured rate can be obtained. 

For direct comparison with data, the detector observable, such as the number of detected photons in data, needs to be converted to the muon energy deposited in the TPC.
The Noble Element Simulation Technique (NEST) model~\cite{NEST2011} is used to convert the number of detected photons in the TPC (the sum of S1 and S2 pulse areas) to deposited energy. 
NEST was tailored for the WS2022 and WS2024 detector configurations to generate conversion factors for each dataset. 
Using the start positions and direction vectors from MUSUN simulations, the tracks of simulated muons were projected towards the TPC.
The positional information for where the muons entered and exited the TPC was used in the NEST simulations that generated both the energy deposition via ionisation and the S1 and S2 signals. 
This relationship between the sum of detected photons in S1 and S2 pulses per event and the corresponding deposited energy, in simulations, was used to convert the number of detected photons in the data into the energy deposition. The energy response is found to be approximately linear below 100 MeV. Above this energy, saturation of PMTs prevents a direct comparison of reconstructed energy in the TPC to that predicted by simulations.

\begin{figure}[!ht]
    \centering
    \begin{subfigure}{.75\textwidth}
        \centering
        \centering\includegraphics[width=\textwidth]{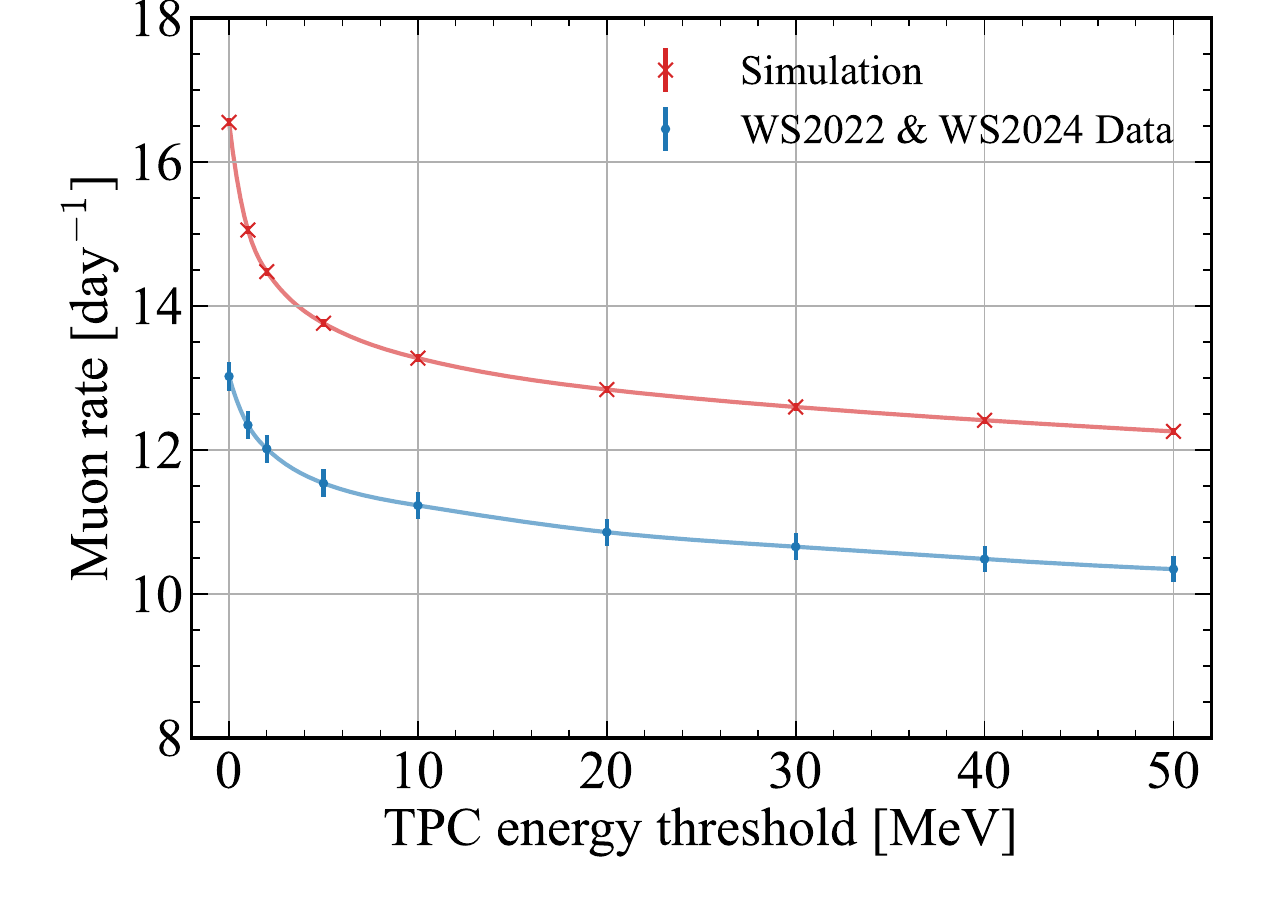}
    \caption{}
    \label{fig:muon_rates}
    \end{subfigure}
    \begin{subfigure}{.75\textwidth}
    \centering
    \centering\includegraphics[width=\textwidth]{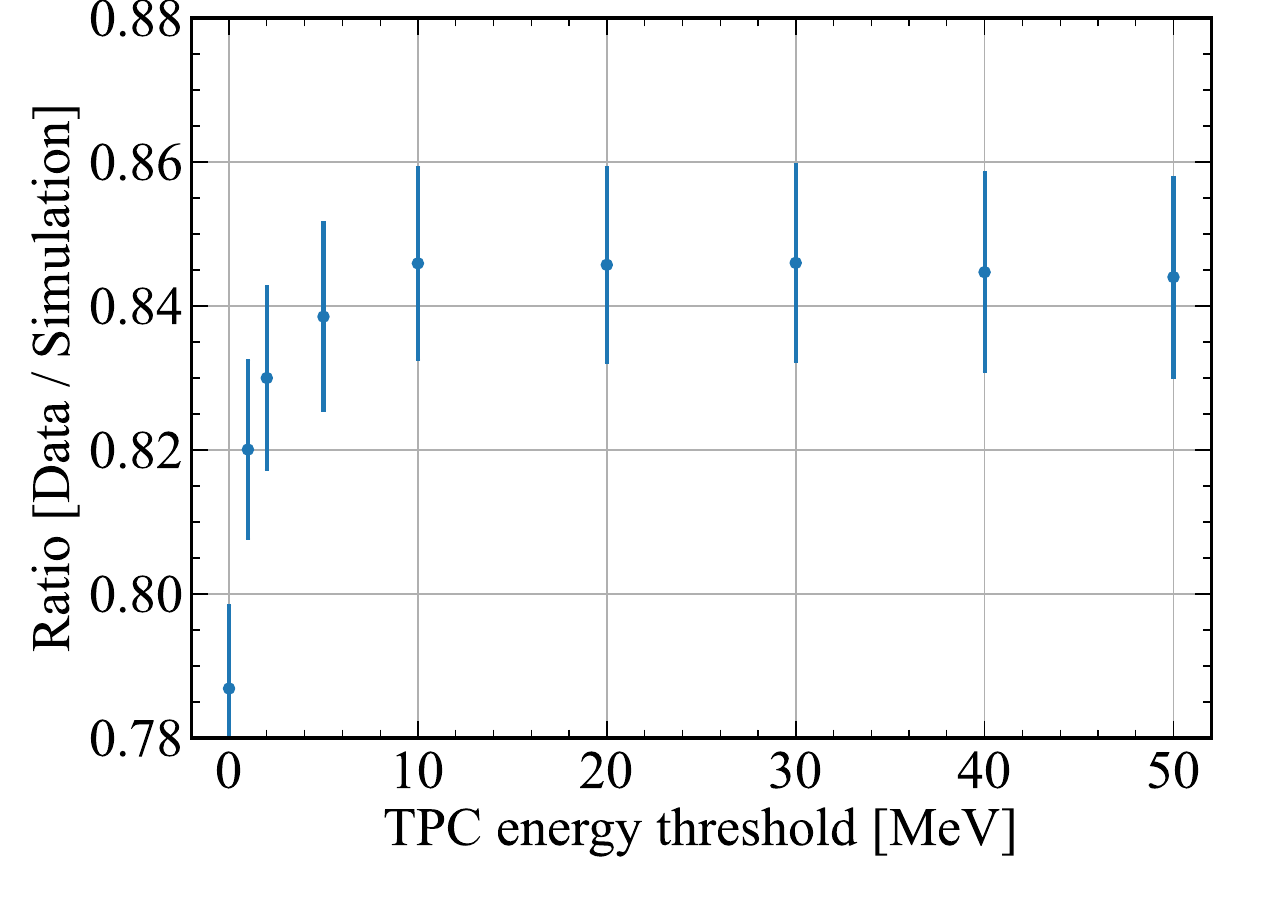}
    \caption{}
    \label{fig:TPCHGLGComp}
    \end{subfigure}
\caption{(a) Muon rate as a function of the energy threshold in the TPC after analysis cuts have been applied to simulations and data. (b) The ratio of data-to-simulated muon event rates as a function of the energy threshold in the TPC.}\label{fig:ratio_rates}
\end{figure}

The average muon rates per day in data and simulations are plotted in \autoref{fig:muon_rates} as a function of the TPC energy threshold. These rates and their ratio\footnote{The ratio is taken to be the number of events in data divided by the number of events in simulations.} are shown in \autoref{tab:results} as a function of the TPC energy threshold.
The ratio of data-to-simulated muon event rates is displayed in \autoref{fig:TPCHGLGComp}, the errors shown are purely statistical from data and simulated samples. As seen from \autoref{tab:results} and \autoref{fig:TPCHGLGComp}, the muon data rate is lower than that from simulations by about 15-20\%. The difference is most likely due to the rock density used in simulations (see \autoref{sec:eval_flux}).

\begin{table}[!ht]
    \centering
    \def\arraystretch{1}%
    \caption{Muon rates from data and simulations. The first column lists the energy thresholds in the TPC. The second column gives the number of events per day that passed the TPC threshold and analysis cuts in the simulations, while the third column provides the corresponding number for data. The fourth column presents the ratio of data to simulated events per day. The first row corresponds to the lowest energy threshold achieved in the analysis.}
    \begin{tabular}{cccc} \hline \hline
    \multicolumn{1}{c}{\textbf{Energy threshold}} & \multicolumn{1}{c}{\textbf{Simulation}} & \multicolumn{1}{c}{\textbf{Data}} & \multicolumn{1}{c}{\textbf{Data/Simulation}} \\
    \multicolumn{1}{c}{\textbf{in the TPC [MeV]}} & \multicolumn{1}{c}{\textbf{rate [day$^{-1}$]}} & \multicolumn{1}{c}{\textbf{rate [day$^{-1}$]}} & \multicolumn{1}{c}{\textbf{ratio}} \\
    \hline
    $0.00144$ & $16.55 \pm 0.04$ & $13.43 \pm 0.19$ & $0.811 \pm 0.012$\\
    $1$ & $15.06 \pm 0.04$ & $12.69 \pm 0.19$ & $0.843 \pm 0.013$\\
    $2$ & $14.48 \pm 0.04$ & $12.27 \pm 0.18$ & $0.847 \pm 0.013$\\
    $5$ & $13.76 \pm 0.04$ & $11.71 \pm 0.18$ & $0.851 \pm 0.014$\\
    $10$ & $13.28 \pm 0.04$ & $11.35 \pm 0.18$ & $0.855 \pm 0.014$\\
    $20$ & $12.84 \pm 0.04$ & $10.94 \pm 0.18$ & $0.852 \pm 0.014$\\
    $30$ & $12.60 \pm 0.04$ & $10.71 \pm 0.17$ & $0.850 \pm 0.014$\\ 
    $40$ & $12.41 \pm 0.04$ & $10.53 \pm 0.17$ & $0.849 \pm 0.014$\\ 
    $50$ & $12.26 \pm 0.04$ & $10.39 \pm 0.17$ & $0.847 \pm 0.014$\\ \hline \hline
    \end{tabular}
    \label{tab:results}
\end{table}

The muon event rate in data and simulations decreases with increasing energy thresholds in the TPC. However, the ratio of the two remains practically constant for thresholds between 5 and 50~MeV with a variation of less than 1~\%. 
High-energy muons typically deposit 5-10~MeV energy per cm in liquid xenon, and practically all muon events with deposited energy above 10~MeV are caused by muons crossing at least a short section of the TPC.
This was confirmed by looking at deposits that were made only by muons and not their secondaries in the simulation. 
Below this energy, the data-to-simulation muon rate ratio varies with the chosen energy threshold because of uncertainties in modelling events where a muon did not pass through the TPC, but its secondaries did (\autoref{tab:results}).
The energy threshold in the TPC was therefore chosen to be 20~MeV. The measured muon daily rate above this threshold is $10.94\pm0.18_\textrm{stat.}$. The uncertainty in the detected-photon-to-energy conversion in the TPC has been used to evaluate the systematic uncertainty for the measured muon rate. Conservatively, we assume that the uncertainty in this conversion factor can be up to a factor of 1.5. 
To account for this, we calculated the muon rates in both data and simulations by varying the energy threshold in both sets by a factor of 1.5. 
This established the uncertainty in the ratio of the data-to-simulation rate to be about 2\%. 

\autoref{tab:cuts} summarises the effect of each analysis cut on the total number of events in data. The time coincidence cut removed the largest percentage of events, 99.84\%. Out of the 338,945,282 events acquired in the WS2022 and WS2024 datasets, our cuts selected 4007 events for the final rate analysis and flux reconstruction.
The combined efficiency of the first two cuts (time coincidence and the OD energy threshold) is $\approx100$\% with a negligible systematic uncertainty (<1\%). This follows from an almost $4\pi$ coverage and 100\% efficiency of the OD and the Skin to detect muon events that give an energy deposition in the TPC. More details about these cuts are given in \autoref{sec:evt_selec}. The last cut and its uncertainty are discussed above. 
Importantly, the same cuts have been applied to simulations.

\begin{table}[!ht]
    \caption{The number of events remaining after each analysis cut is applied sequentially to the WS2022 and WS2024 data. The ``timing coincidence" cut has been defined in Section~\ref{sec:evt_selec}.}
    \centering
    \def\arraystretch{1.2}
    \begin{tabular}{cc}
    \hline\hline
       \textbf{Cuts}&\textbf{No. of Events} \\
       \hline
       None & $3.389\times10^8$ \\
       Timing Coincidence & $5.327\times10^5$ \\
       OD Energy $>8$~MeV & $4.873\times10^3$ \\
       TPC Energy $>20$~MeV & $4.007\times10^3$ \\
       \hline\hline
    \end{tabular}
    \label{tab:cuts}
\end{table}

Liquid scintillators have a typical light yield of a few thousand photons per MeV. The OD is viewed by 120 8-inch PMTs located on a support structure more than a meter away from the outer surface of the OD, compared to 494 3-inch PMTs looking at the TPC. 1~MeV energy deposition in the OD produces about 250 photoelectrons.
The OD has been calibrated with radioactive sources at keV and MeV energies in LZ, however, the obtained calibration factor is not applicable at muon energy depositions, and thus a dedicated tuning of the energy scale is carried out in this work.

\begin{figure}[!ht]
    \centering
    \includegraphics[width=0.9\textwidth]{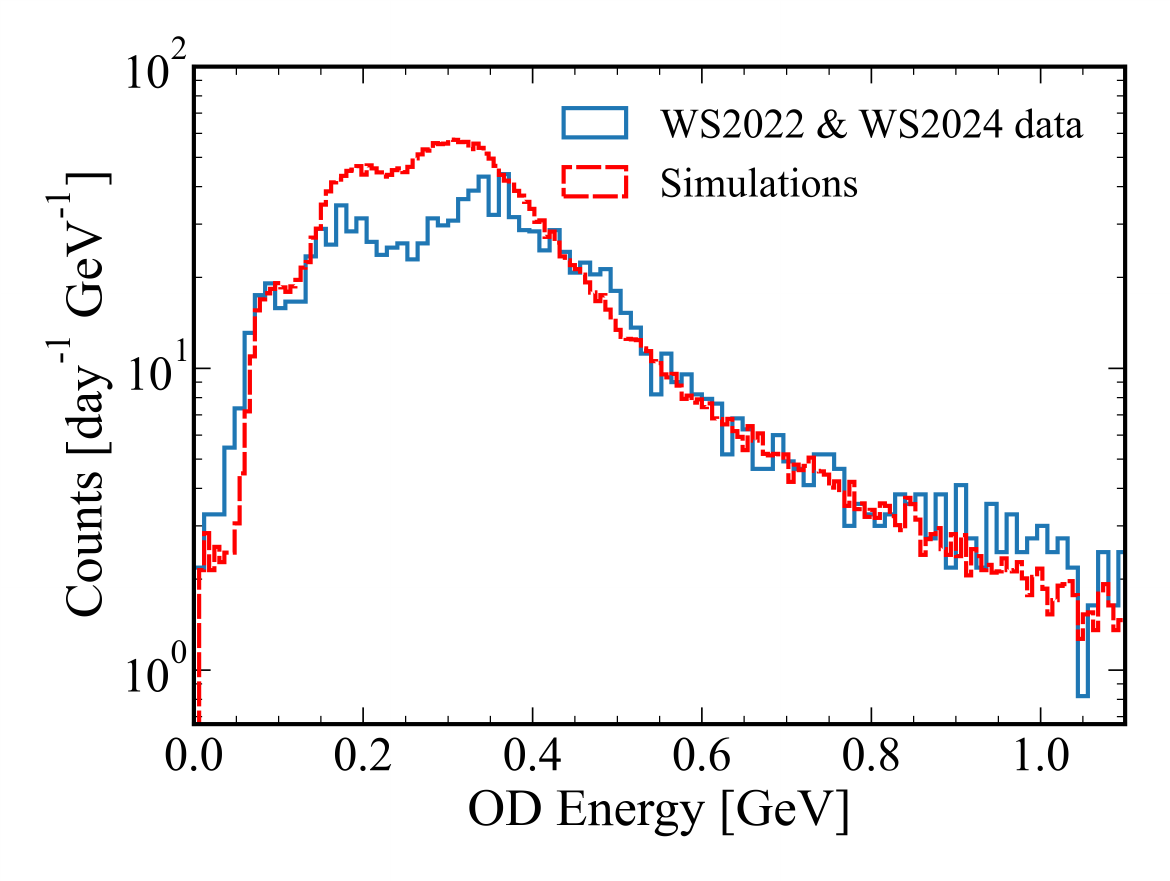}
    \caption{Spectra of energy depositions from muon events in the OD in data and simulations. Energy deposited in the OD in data was obtained from the total pulse area using a single scaling factor to match the simulated spectrum. The shape of the spectra is due to the path dependence of the muon through the complex OD tanks geometry.}
    \label{fig:OD_comp}
\end{figure}

The comparison of measured and simulated energy deposition spectra in the OD is shown in \autoref{fig:OD_comp}. 
We have not simulated light emission, transport, and detection in the OD due to the high number of emitted scintillation photons and limitations in computation power. 
Therefore, while a perfect match between the two spectra is not expected, the agreement in their shapes is reasonably good, with a lower rate of measured events compared to simulation.
Non-uniformity of the light collection in the OD and complexity of the muon pulse reconstruction by the LZ software currently tuned to low-energy events could be reasons for a difference between the shapes of the two spectra.

Although muons can reach the LZ depth as single muons and in bundles (multiple muons produced in the same air shower), the fraction of multiple muons at this depth is relatively small. Different experiments reported fractions of multiple muons relative to single muons of about 7\% at 2.1~km~w.~e.~\cite{soudan2}, 15\% at 3~km~w.~e.~\cite{macro}, 4\% at 4~km~w.~e.~\cite{utah}, and 1.3\% at 6~km~w.~e.~\cite{kgf}. The fraction of multiple muons depends on the detector depth, zenith angle, and size, leading to the observed variations. For all reported measurements, the fraction of muon pairs separated by less than 1.5~m (the diameter of the LZ TPC) is smaller than 5\%. This gives the fraction of muons to be detected within 1.5~m distance from another muon as $<1\%$ relative to the number of single muons. Ref.~\cite{utah2} measured the fraction of muon pairs within a 1~m$^2$ area as about 0.001 relative to the number of single muons. This agrees with the results of modelling~\cite{gaisser-stanev}. We conclude then that with the current selection criteria, the fraction of multiple muons to be detected by LZ is less than 1\% relative to the number of single muons, which is smaller than the uncertainty of the measurements and can be neglected.

Given our overall statistical uncertainty of about 1.6\%, our measurements cannot reveal annual modulation of the muon flux expected to have an amplitude of less than 2\% (see, for example, Ref.~\cite{Borexino}). Our measurements cover almost the whole year period and hence the results refer to the muon rate and flux close to the average over the year.

\section{Evaluation of the Muon Flux}\label{sec:eval_flux}
In this work, we measured the rate of muons passing through the three detectors of LZ. A useful parameter that can be extracted from this measurement is the total muon flux. Below, we describe the procedure of muon flux derivation from the LZ data and compare this flux with other measurements. We also discuss the most likely reason for the difference between LZ data and our initial muon model.

The measured rate can be converted to a muon flux by scaling the simulated muon flux by the same ratio as that for the measured-to-simulated rate:
\begin{equation}
    F_{m} = F_{s}\times\frac{R_m}{R_s},
\label{eq:flux}
\end{equation}
where $F_{m}$ and $F_{s}$ are the measured (evaluated from the measured rate) and simulated muon fluxes through the surface of a sphere (unit detection efficiency at all angles), respectively, and $R_{m}$ and $R_{s}$ are the measured and simulated muon rates through the detector.

This scaling is based on the common inputs to our muon model, namely, muon energy spectra and angular distributions at SURF. It also assumes that muon transport through the detector and detector response are simulated correctly in BACCARAT. Simulation of muon-induced cascades and their development in and outside the detector has associated uncertainties that are difficult to estimate. By requiring the energy deposition in the TPC to be greater than 20~MeV, we have effectively selected events when a muon passes through the TPC, removing a relatively small contribution of events 
when a muon minimally clips off the TPC edge or only low-energy 
secondary particles enter the TPC. No other energy cut for the TPC was included; therefore, all events with a muon in the TPC, Skin and OD passing the energy thresholds (both in data and simulations) are considered in the analysis.

The simulation of the muon track passing through the detector and muon ionisation loss along the track is known to be handled accurately by GEANT4. The detector response model has been validated by comparing various backgrounds in data and simulations at keV-MeV energies (see, for instance, Refs.~\cite{LZfirstResults,WS2024,LZ_bkg}). Moreover, an accurate simulation of detector response becomes non-critical because the rate of events does not include information about energy deposition apart from the energy threshold.

The ratio of measured-to-simulated muon rates in \autoref{tab:results} remains constant within statistical uncertainty for TPC energy thresholds of 5-50~MeV. For a TPC threshold of 20~MeV, the measured-to-simulated muon rate ratio is $0.852\pm0.014_\textrm{stat.}\pm0.017_\textrm{sys.}$. The systematic uncertainty associated with the TPC energy conversion is estimated by varying the threshold asymmetrically in data and simulations: thresholds in data are increased (decreased) while those in simulations are decreased (increased) by a factor of 1.5. The resulting rate ratios are compared to the nominal value at 20~MeV, and the largest difference is taken as the systematic uncertainty.

Using \autoref{eq:flux}, the data-to-simulations rate ratio of $0.852\pm0.014_\textrm{stat.}\pm0.017_\textrm{sys.}$, and the simulated muon flux of $6.16\times10^{-9}\textrm{cm}^{-2}\textrm{s}^{-1}$, we derive the value for the muon flux from the rate measurements as $(5.25\pm0.09_\textrm{stat.}\pm0.10_\textrm{sys.})\times10^{-9}~\textrm{cm}^{-2}\textrm{s}^{-1}$. 

This total muon flux is for the surface of the box where muons were originally sampled in the simulations, namely 7~m into rock above the cavern for near vertical muons (the mean zenith angle of muons arriving at the LZ detector is $27^{\circ}$). The flux depends on the position in and around the cavern. Our simulations show that the muon flux at the top surface of the cavern is about 3\% lower than at 7~m into rock and is $(5.09 \pm0.08_\textrm{stat.}\pm0.10_\textrm{sys.})\times10^{-9}\textrm{cm}^{-2}\textrm{s}^{-1}$. This flux is consistent with the measurement of muons by the M\footnotesize{AJORANA} \normalsize{D}\footnotesize{EMONSTRATOR }\normalsize in the nearby cavern reported in Ref.~\cite{majorana} ($(5.31\pm0.17_\textrm{stat.})\times10^{-9}~\textrm{cm}^{-2}\textrm{s}^{-1}$).

The low muon rate at SURF and the complexity of track reconstruction with the optical readout (PMTs) did not allow us to study the muon angular
distribution.

The difference between the measured and simulated muon fluxes can be attributed to a lower rock density used in simulations. We assume here that the uncertainties in other inputs should not largely affect the muon flux. The surface profile and laboratory position are known with sufficient accuracy. The parametrisation of the muon energy spectrum and angular distribution at the Earth's surface~\cite{Gaisser_1990} in our model is based on experimental data, and the dependence of the muon flux on the distance travelled by muons in known rocks was validated against the accurate measurements by the LVD detector at LNGS~\cite{LVD:1998lir} and other labs~\cite{musun}. 

By attributing the lower measured muon flux to a higher average rock density, we can estimate the realistic average rock density above and around the LZ location. We calculated the muon flux at SURF using different rock densities above the laboratory to match the reconstructed measurement. For simplicity, a `flat' surface profile was assumed in these simulations only, which accounted for Earth's curvature, but not for any features in the surface profile. The simulated flux matches the measured one with an average rock density of $(2.77\pm0.01)~\textrm{g~cm}^{-3}$, which is 2.5\% higher than the initial muon model assumption, but lower than that reported in Ref.~\cite{majorana} ($(2.89\pm0.06)~\textrm{g~cm}^{-3}$). Our density is lower than the measured value from Ref.~\cite{heise} ($2.85~\textrm{g~cm}^{-3}$) as an average for a $45^{\circ}$ cone above the LZ location, but is still in agreement within a 4\% uncertainty in the overburden quoted in that paper.

The precise rock composition around and above the laboratory is not known, and the rock is not uniform. Hence, choosing a different rock composition, for instance, the rock Yates amphibolite~\cite{Caddey} present around the laboratory (with $\langle Z\rangle=12.6$ and $\langle A\rangle\:=25.3$), results in a slightly different evaluated density from our muon rate measurements, namely: $(2.75\pm0.01)~\textrm{g~cm}^{-3}$. The average of the two values can be taken as an estimate of the average rock density and the difference between the two density values and their average can be taken as an additional systematic uncertainty, resulting in a density of $(2.76\pm0.01_\textrm{stat.}\pm0.01_\textrm{sys.})~\textrm{g~cm}^{-3}$.

\section{Conclusions}\label{sec:conc}
The LZ experiment has measured the rate of muons passing through the three detectors: TPC, Skin, and OD, as $10.94\pm0.18_\textrm{stat.}~\textrm{day}^{-1}$ with energy thresholds of 20~MeV in the TPC and 8~MeV in the OD. This rate is lower than calculated from the initial muon model by a factor of $0.852\pm0.014_\textrm{stat.}\pm0.017_\textrm{sys.}$. By scaling the simulated muon flux by the ratio of data-to-simulated rates and extrapolating the flux to the top of the Davis cavern we derived the muon flux from the measurements as $(5.09\pm0.08_\textrm{stat.}\pm0.10_\textrm{sys.})\times10^{-9}~\textrm{cm}^{-2}\textrm{s}^{-1}$. Assuming that the difference between the measured and calculated muon rates is due primarily to an unduly low rock density used in the muon model and that the density is uniform above and around the detector, the corrected rock density has been found as $(2.76\pm0.01_\textrm{stat.}\pm0.01_\textrm{sys.})~\textrm{g~cm}^{-3}$, 2.5\% higher than in the initial model.

\section*{Acknowledgements}
\sloppy{
We would like to thank Dr Jaret Heise, the Science Director of SDSTA, for useful discussions and for providing us with the composition of the Yates amphibolite rock. The research supporting this work took place in part at the Sanford Underground Research Facility (SURF) in Lead, South Dakota.
Funding for this work is supported by the U.S. Department of Energy, Office of Science, Office of High Energy Physics under Contract Numbers DE-AC02-05CH11231, DE-SC0020216, DE-SC0012704, DE-SC0010010, DE-AC02-07CH11359, DE-SC0015910, DE-SC0014223, DE-SC0010813, DE-SC0009999, DE-NA0003180, DE-SC0011702, DE-SC0010072, DE-SC0006605, DE-SC0008475, DE-SC0019193, DE-FG02-10ER46709, UW PRJ82AJ, DE-SC0013542, DE-AC02-76SF00515, DE-SC0018982, DE-SC0019066, DE-SC0015535, DE-SC0019319, DE-SC0024225, DE-SC0024114, DE-AC52-07NA27344 and DE-SC0012447.
This research was also supported by U.S. National Science Foundation (NSF); the UKRI's Science and Technology Facilities Council (STFC) under award numbers ST/W000490/1, ST/W000482/1, ST/W000636/1, ST/W000466/1, ST/W000628/1, ST/W000555/1, ST/W000547/1, ST/W00058X/1, ST/ X508263/1, ST/V506862/1, ST/X508561/1, ST/V507040/1, ST/W507787/1, ST/R00318-1/1, ST/R003181/2,  ST/W507957/1, ST/X005984/1, ST/X006050/1; Portuguese Foundation for Science and Technology (FCT) under award numbers PTDC/FIS-PAR/2831/2020; the Institute for Basic Science, Korea (budget number IBS-R016-D1); the Swiss National Science Foundation (SNSF)  under award number 10001549. This research was supported by the Australian Government through the Australian Research Council Centre of Excellence for Dark Matter Particle Physics under award number CE200100008. We acknowledge additional support from the UK Science and Technology Facilities Council (STFC) for PhD studentships and the STFC Boulby Underground Laboratory in the U.K, the GridPP \cite{faulkner2005gridpp,britton2009gridpp} and IRIS Collaborations, in particular at Imperial College London and additional support by the University College London (UCL) Cosmoparticle Initiative, and the University of Zurich. We acknowledge additional support from the Center for the Fundamental Physics of the Universe, Brown University. K.T. Lesko acknowledges the support of Brasenose College and Oxford University. The LZ Collaboration acknowledges the key contributions of Dr. Sidney Cahn, Yale University, in the production of calibration sources. This research used resources of the National Energy Research Scientific Computing Center, a DOE Office of Science User Facility supported by the Office of Science of the U.S. Department of Energy under Contract No. DE-AC02-05CH11231. We gratefully acknowledge support from GitLab through its GitLab for Education Program. The University of Edinburgh is a charitable body, registered in Scotland, with the registration number SC005336. The assistance of SURF and its personnel in providing physical access and general logistical and technical support is acknowledged. We acknowledge the South Dakota Governor's office, the South Dakota Community Foundation, the South Dakota State University Foundation, and the University of South Dakota Foundation for use of xenon. We also acknowledge the University of Alabama for providing xenon. For the purpose of open access, the authors have applied a Creative Commons Attribution (CC BY) license to any Author Accepted Manuscript version arising from this submission. Finally, we respectfully acknowledge that we are on the traditional land of Indigenous American peoples and honour their rich cultural heritage and enduring contributions. Their deep connection to this land and their resilience and wisdom continue to inspire and enrich our community. We commit to learning from and supporting their effort as original stewards of this land and to preserve their cultures and rights for a more inclusive and sustainable future. 
}

\end{document}